\documentclass[reqno]{conm-p-l}
\usepackage{amssymb,amsmath,amsthm}

\newtheorem{thm}{Theorem}
\newtheorem{cor}{Corollary}
\newtheorem{lem}{Lemma}

\theoremstyle{definition}

\theoremstyle{remark}
\newtheorem{rem}{Remark}

\numberwithin{equation}{section}

\newcommand{\K}{{\mathcal K}}
\newcommand{\X}{{\bf X}}
\newcommand{\Enqm}{E^{\rm QM}}
\newcommand{\pgp}{\phi^{\rm GP}}
\newcommand{\widehatpgp}{{\widehat \phi}^{\rm GP}}
\newcommand{\kk}{{\bf k}}
\newcommand{\rmax}{\rho_{\rm max}}
\newcommand{\rmin}{\rho_{\rm min}}
\newcommand{\al}{\alpha}

\newcommand{\R}{{\mathbb R}}

\newcommand{\eps}{\varepsilon}

\newcommand{\x}{{\bf x}}

\newcommand{\Tr}{{\rm Tr}}
\newcommand{\half}{\mbox{$\frac{1}{2}$}}
\newcommand{\E}{{\mathcal E}^{\rm GP}}
\newcommand{\Engp}{E^{\rm GP}}

\begin{document}

\title{Bose-Einstein Condensation of Dilute Gases in Traps}
\author{Elliott H. Lieb}
\address{Departments of Mathematics and 
Physics, Jadwin Hall, Princeton University,
P.O. Box 708, Princeton, New Jersey 08544, USA}
\email{lieb@math.princeton.edu}
\thanks{\copyright\ 2002 by the authors. Reproduction of this work in its entirety, by any means, is permitted for non-commercial purposes.}
\thanks{The first author was supported in part 
by NSF grant PHY 01 39984.}
\author{Robert Seiringer}
\address{Department of Physics, Jadwin Hall, Princeton University,
P.O. Box 708, Princeton, New Jersey 08544, USA. On leave
from Institut f\"ur Theoretische Physik, Universit\"at Wien,
Boltzmann\-gasse 5, 1090 Vienna, Austria.}
\email{rseiring@math.princeton.edu}
\thanks{The second author was supported by the Austrian Science Fund in the form of an Erwin
Schr\"odinger Fellowship.}

\subjclass[2000]{81V70, 35Q55, 46N50}
\date{June 17, 2002}

\begin{abstract}
The ground state of a gas of Bosons confined in an external trap
potential and interacting via repulsive two-body forces has recently
been shown to exhibit complete Bose-Einstein condensation in the dilute
limit, yielding for the first time a rigorous proof of this phenomenon
in a physically realistic setting. We give here an account of this
work about the Gross-Pitaevskii limit where the particle
number $N$ goes to infinity with $Na$ fixed, where $a$ is the
scattering length of the interparticle interaction, measured in units
of the trap size.
\end{abstract}

\maketitle

\section{Introduction and Main Results}

During the last few years it has become experimentally feasible
to realize the long-predicted Bose-Einstein condensation (BEC) of
gases by confining them in traps at very low temperatures. A rigorous
theoretical demonstration of this phenomenon -- starting from the
basic many-body Hamiltonian of interacting particles -- is, however, a
very difficult task. Following
\cite{LS02}, we will provide in this paper such a rigorous
derivation for the ground state of Bosons in an external trap
interacting with repulsive pair potentials, and in the well-defined
limit in which the Gross-Pitaevskii (GP) formula is applicable. It is
the first proof of BEC for interacting particles in a continuum (as
distinct from lattice) model and in a physically realistic
situation. 
We present here a detailed version of the proof given in
\cite{LS02}. We remark that an extension
of the results presented here was recently obtained in \cite{lsy02a}, where
it was shown that the ground state is 100\% superfluid in the GP
limit.

The Gross-Pitaevskii limit under discussion here is a mathematically
simpler limit than the usual thermodynamic limit in which the average
density is held fixed as the particle number goes to infinity. 
A proof of BEC in this
limit is still missing. The only available rigorous results
concern the non-interacting gas, various mean-field and other toy
models (see, e.g., \cite{lauw} for a recent preprint), or the hard
core lattice gas \cite{LKS}.  

In the GP limit one also lets the range
of the potential go to zero as $N$ goes to infinity, but in such a way
that the overall effect is non-trivial. That is, the combined effect
of the infinite particle limit and the zero range limit is such as to
leave a measurable residue --- the GP function.  The limit in which
the GP function can be expected to be equal to the condensate wave
function should be chosen so that {\it all three} terms in the GP
functional (\ref{defgp}) make a contribution. This indicates that
fixing $Na$ as $N\to\infty$ is the right thing to do, and this is
quite relevant since experimentally $N$ can be quite large, $10^6$ and
more, and $Na$ can range from about 1 to $10^4$ \cite{PS}.  Fixing
$Na$ also means that we really are dealing with a dilute system,
because the mean density $\bar \rho$ is then of the order $N$ and
hence $a^3\bar\rho\sim N^{-2}$, meaning that the range of the
interaction is much shorter than the mean particle distance.

We shall now describe the setting more precisely. Let $H$ be the
Hamiltonian for $N$ identical Bosons in a trap potential $V$ in
$\R^3$, interacting via a pair potential~$v$:
\begin{equation}\label{ham}
H=\sum_{i=1}^N \big(-\Delta_i+V(\x_i)\big)+\sum_{1\leq i<j\leq
N} v(\x_i-\x_j) \ .
\end{equation}
It acts on the subspace of totally symmetric functions in
$\bigotimes^N L^2(\R^3)$. Units are chosen such that $\hbar=2m=1$,
where $m$ denotes the particle mass.  We assume the trap potential $V$
to be a locally bounded function that tends to infinity as
$|\x|\to\infty$. The interaction potential $v$ is assumed to be
nonnegative, spherically symmetric, and have a finite range $R_0$. Its
scattering length will be denoted by $a$. (For the definition of
scattering length, see
\cite{LSY00} or \cite{LY01}.) The finiteness of the
range will be assumed for simplicity, but this assumption can be
relaxed. Note that we do not demand $v$ to be locally integrable; it
is allowed to have a hard core, which forces the wave functions to
vanish whenever two particles are close together. In the following, we
want to let $a$ vary with $N$, and we do this by scaling, i.e., we
write $v(\x)=v_1(\x/a)/a^2$, where $v_1$ has scattering length $1$,
and keep $v_1$ fixed when varying $a$. The ground state energy of $H$,
denoted by $\Enqm$, then depends only on $N$ and $a$ (for fixed $V$
and $v_1$), so the notation $\Enqm(N,a)$  is justified. The
results below are independent of the particular shape of the
interaction potential $v_1$.

The Gross-Pitaevskii functional is given by
\begin{equation}\label{defgp}
\E[\phi]=\int\big(|\nabla\phi(\x)|^2+V(\x) |\phi(\x)|^2 + g
|\phi(\x)|^4\big)d^3\x \ .
\end{equation}
Here $g$ is a positive parameter that is related to the particle
number and the scattering length of the interaction potential
appearing in (\ref{ham}) via
\begin{equation}
 g=4\pi N a \ .
\end{equation}
We denote
by $\pgp$  the minimizer of $\E$ under the normalization condition
$\int|\phi|^2=1$. Existence, uniqueness, and  some regularity
properties of $\pgp$ were proved in the appendix of \cite{LSY00}. In
particular, $\pgp$ is continuously differentiable and strictly
positive. Of course $\pgp$ depends on $g$, but we omit this dependence
for  simplicity of  notation.  For later use, we define the projector
\begin{equation}
P^{\rm GP}= |\pgp\rangle\langle \pgp|\ .
\end{equation}
The minimizer $\pgp$ fulfills the variational equation
\begin{equation}\label{GPeq}
-\Delta \pgp + V \pgp + 2 g (\pgp)^3=\mu^{\rm GP} \pgp \ ,
\end{equation}
which is called the {\it GP equation}. Here $\mu^{\rm GP}$ is the
chemical potential, given by
\begin{equation}\label{mugp}
\mu^{\rm GP} = \Engp + g \int |\pgp|^4 \ ,
\end{equation}
with $\Engp=\Engp(g)$ the lowest energy of $\E$ under the condition
$\int |\phi|^2=1$.

It was shown in \cite{LSY00} (see Theorem \ref{T2} below) that, for each
fixed $g$, the minimization of the GP functional correctly reproduces
the large $N$ asymptotics of the ground state energy and density of
$H$ -- but no assertion about BEC in this limit was made. This was
extended in \cite{LS02}, where complete BEC was proved. The precise
statement is as follows.

Let $\Psi$ denote the nonnegative and normalized ground state of
$H$. BEC refers to the reduced one-particle density matrix
\begin{equation}
\gamma(\x,\x')=N\int \Psi(\x,\X) \Psi(\x',\X) d\X \ ,
\end{equation} 
where we denoted $\X=(\x_2,\dots,\x_N)$ and $d\X=\prod_{j= 2}^N
d^3\x_j$ for short. It is the kernel of a positive trace class operator
$\gamma$ on $L^2(\R^3)$, with $\Tr[\gamma]=N$.

Complete (or 100\%) BEC is defined to be the property that
$\mbox{$\frac{1}{N}$}\gamma(\x,\x')$ becomes a simple product
$\varphi(\x)\varphi(\x')$ as $N\to \infty$, in which case $\varphi$ is
called the {\it condensate wave function}. I.e.,
$\mbox{$\frac{1}{N}$}\gamma$ converges to a one-dimensional
projection. In particular, there is one eigenvalue of $\gamma$ of the
order $N$ (even equal to $N$ in the limit), and all the others are of
lower order. In the GP limit, i.e., $N\to\infty$ with $g=4\pi N a$
fixed, we can show that this is the case, and the condensate wave
function is, in fact, the GP minimizer $\pgp$.

\begin{thm}[{\bf Bose-Einstein Condensation}]\label{T1}
For each fixed $g$
\begin{equation}
\lim_{N\to\infty} \frac 1 N \gamma(\x, \x') =
\pgp(\x)\pgp(\x') 
\end{equation}
in trace class norm, i.e., 
$\Tr \big[|\gamma/N - P^{\rm GP}|\big] \to 0$.
\end{thm}

\begin{rem} Theorem \ref{T1} implies that there is 100\%
condensation for all $n$-particle reduced density matrices of $\Psi$,
i.e., they converge to the one-dimensional projector onto the
corresponding $n$-fold product of $\pgp$. See \cite{LS02} for details.
\end{rem}

\begin{rem}
Convergence does not hold
for the Sobolev norm $\Tr[|(1-\Delta)(\gamma/N - P^{\rm GP})|]$. 
It can be shown \cite{LS02} that
\begin{equation}\label{1.9}
\lim_{N\to\infty}\frac 1N \Tr[-\Delta\gamma]= \Tr[-\Delta P^{\rm GP}] + g s \int|\pgp|^4
\end{equation}
for some parameter $0<s\leq 1$ depending only on the interaction
potential $v_1$.
\end{rem}

A corollary of Theorem \ref{T1}, important for the interpretation of
experiments, concerns the momentum distribution of the ground
state.

\begin{cor}[\textbf{Convergence of momentum distribution}]\label{C1} 
Let
\begin{equation}
\widehat\rho (\kk)= \int\gamma(\x, \x') \exp [i \kk\cdot (\x
-\x')]
 d^3\x d^3\x'
\end{equation}
denote the one-particle momentum  density of $\Psi$. Then, for
fixed $g$,
\begin{equation}
 \lim_{N\to\infty} \frac 1N
\widehat\rho(\kk)=|\widehatpgp(\kk)|^2
\end{equation} 
strongly in
$L^1(\R^3)$. Here, $\widehatpgp$ denotes the Fourier
transform of $\pgp$.
\end{cor}

\begin{proof} 
If ${\mathcal F}$ denotes the (unitary) operator `Fourier
transform' and if $h$ is an arbitrary $L^\infty$-function,
then
\begin{equation}
\left|\frac 1N\int \widehat\rho h-\int |\widehatpgp|^2 h\right|
=\left|\Tr[{\mathcal F}^{-1}
(\gamma/N-P^{\rm GP}){\mathcal F}h]\right|
 \leq \|h\|_\infty \Tr \big[|\gamma/N-P^{\rm GP}|\big] \ ,
\end{equation}
from which we conclude that
\begin{equation}
\|\widehat\rho/N-|\widehatpgp|^2 \|_1\leq \Tr\big[|\gamma/N-P^{\rm GP}|\big]\ .
\end{equation}
\end{proof}

It is important to note that in the limit considered $v$ becomes a
{\em hard} potential of {\em short range}. This is the opposite of the
usual mean field limit, where the strength of the potential goes to
zero while its range tends to infinity.

We also wish to emphasize that in this GP limit the fact that there is
100\% condensation does not mean that no significant interactions
occur.  The kinetic and potential energies can differ markedly from
those obtained with a simple variational function that is an $N$-fold
product of one-body condensate wave functions. This assertion might
seem paradoxical, and the explanation is that near the GP limit the
region in which the wave function differs from the condensate function
has a tiny volume that goes to zero as $N\to \infty$. Nevertheless,
the interaction energy, which is proportional to $N$, resides in this
tiny volume.

Before proving Theorem \ref{T1}, let us state some prior results on which we
shall build. These results concern the asymptotic behavior of the
ground state energy and density of (\ref{ham}) in the limit
$N\to\infty$ with $g=4\pi N a$ fixed. The following Theorem \ref{T2} was
proved in \cite{LSY00}.

\begin{thm}[{\bf Asymptotics of Energy and Density}]\label{T2}
Let
$\rho(\x)=\gamma(\x,\x)$ denote the density of the ground state of
$H$. For fixed $g=4\pi Na$,
\begin{equation}\label{part1a}
\lim_{N\to\infty}\frac 1N \Enqm(N,a)=\Engp(g)
\end{equation}
and
\begin{equation}\label{part1}
\lim_{N\to\infty}\frac 1N \rho(\x) = |\pgp(\x)|^2
\end{equation}
in the sense of weak convergence in $L^1(\R^3)$. 
\end{thm}

\begin{rem} The convergence in (\ref{part1}) was  shown in
\cite{LSY00} to be in the weak $L^1(\R^3)$ sense, but our result
here implies strong convergence, in fact. This can easily be deduced
from Theorem \ref{T1} (cf. the proof of Corollary~\ref{C1}).
\end{rem}

We now give a brief outline of the proof of Theorem
\ref{T1}. There are two essential ingredients. The first is a proof
that the part of the kinetic energy that is associated with the
interaction $v$ is mostly located in small balls surrounding each
particle. More precisely, these balls can be taken to have radius
$N^{-7/17}$, which is much smaller than the mean particle spacing
$N^{-1/3}$. This allows us to conclude that the function of $\x$
defined for each fixed value of $\X$ by
\begin{equation}\label{deff}
f_\X(\x)=\frac 1{\pgp(\x)} \Psi(\x,\X)\geq 0
\end{equation}
has the property that $\nabla_\x f_\X(\x)$ is almost zero
outside the small balls centered at points of $\X$.
This is made precise in Lemma \ref{L1}. 

The complement of the small balls has a large volume but it can be
a weird set; it need not even be connected. Therefore, the
smallness of $\nabla_\x f_\X(\x)$ in this set does not guarantee
that $f_\X(\x)$ is nearly constant (in $\x$), or even that it is
continuous. We need $f_\X(\x)$ to be nearly constant in order to
conclude BEC. What saves the day is the knowledge that the total
kinetic energy of $f_\X(\x)$ (including the balls) is not huge.
The result that allows us to combine these two pieces of
information in order to deduce the almost constancy of $f_\X(\x)$
is the generalized Poincar\'e inequality in Lemma \ref{L2}. 

\section{The Proof}

We start with the following Lemma. It shows that to leading order all
the interaction energy is concentrated in small balls. The proof we
give here is a detailed version of the one sketched in
\cite{LS02}. Instead of referring to the methods of \cite{LSY00}, we
follow a slightly different route and use the (partly simpler) methods
of the review article \cite{lssy}.

Throughout the paper we suppress the dependence on $N$ for simplicity of notation. For instance, in Eq. (\ref{neu}) both $\X$ and $f_\X$ depend on $N$. 

\begin{lem}[{\bf Localization of the Energy}] \label{L1}
For fixed $\X$ let
\begin{equation}\label{defomega} \Omega_\X=\left\{\x\in \R^3
\left| \, \min_{k\geq 2}|\x-\x_k|\geq
N^{-7/17}\right\}\right. \ .
\end{equation}  
Then 
\begin{equation}\label{neu} 
\lim_{N\to\infty}
\int d\X \int_{\Omega_\X} d^3\x |\pgp(\x)|^2 |\nabla_\x
f_\X(\x)|^2 = 0 \ .
\end{equation}
\end{lem}

Note that (\ref{neu}) does not imply that the kinetic energy of $f_\X$
goes to zero. In fact, it can be shown \cite{LS02,CS} that if one replaces
$\Omega_\X$ by $\R^3$ in (\ref{neu}) the result is
\begin{equation}\label{bound}
\lim_{N\to\infty} \int d\X \int_{\R^3} d^3\x |\pgp(\x)|^2 
|\nabla_\x f_\X(\x)|^2 = gs\int |\pgp(\x)|^4 d^3\x 
\end{equation}
(compare with (\ref{1.9})). 
Here $0<s\leq 1$ is a parameter depending only on the interaction
potential $v$. (For example, $s=1$ in the case of hard core Bosons; in
general, $s<1$.) The right side of (\ref{bound}) is $O(1)$ in the
limit considered, and is not necessarily small. Thus Lemma~\ref{L1}
shows that all the interaction energy is localized in small balls of
radius $N^{-7/17}$ surrounding each particle.

\begin{proof}[Proof of Lemma \ref{L1}]
We shall show that
\begin{multline} \label{lowbound}
\int d\X \int_{\Omega_\X^c} d^3\x\, |\pgp(\x)|^2 |\nabla_\x
f_\X(\x)|^2\\ +\int d\X \int d^3\x\, |\pgp(\x)|^2 |f_\X(\x)|^2
\left[ \half\sum_{k\geq 2} v(\x-\x_k)  - 2 g |\pgp(\x)|^2 \right] \\ 
\geq -g \int|\pgp(\x)|^4 d^3\x - o(1)
\end{multline} 
as $N\to \infty$. Here $\Omega_\X^c$ is the complement of $\Omega_\X$
in $\R^3$. We claim that this implies the assertion of the Lemma. To
see this, note that the left side of (\ref{lowbound}) can be written
as
\begin{equation}
\frac 1N E^{\rm QM}-\mu^{\rm GP} - \int
d\X \int_{\Omega_\X} d^3\x\, |\pgp(\x)|^2 |\nabla_\x
f_\X(\x)|^2 \ ,
\end{equation}
where we used partial integration and the GP equation (\ref{GPeq}),
and also the symmetry of $\Psi$. The convergence of the energies in Theorem
\ref{T2} and the relation (\ref{mugp}) now imply the desired result.

We are left with the proof of (\ref{lowbound}). This is actually just
a detailed examination of the lower bounds to the energy derived in
\cite{LSY00} and \cite{LY98}, and we use the methods from there.
 The proof presented here follows closely \cite{lssy} and differs
 slightly from \cite{LSY00}.

Writing $f_\X(\x)=\Pi_{k\geq 2}\pgp(\x_k)F(\x,\X)$ and using that $F$
is symmetric in the particle coordinates, we see that (\ref{lowbound})
is equivalent to
\begin{equation}\label{qf} 
\frac 1N Q(F)\geq -g
\int|\pgp|^4 - o(1)\ , 
\end{equation} 
where $Q$ is the quadratic form
\begin{multline}\label{qf2}
Q(F)=\sum_{i=1}^{N} \int_{\Omega_i^c} |\nabla_i
F|^2\prod_{k=1}^{N}|\pgp(\x_k)|^2d^3\x_k\\ 
+\sum_{i=1}^N  \int \left[
\sum_{j\neq i} v(\x_i-\x_j) -2g 
|\pgp(\x_i)|^2 \right] |F|^2\prod_{k=1}^{N}|\pgp(\x_k)|^2d^3\x_k \ ,
\end{multline} 
with
$\Omega_i^c=\{(\x_1,\X)\in\R^{3N}| \, \min_{k\neq i}|\x_i-\x_k|\leq
N^{-7/17}\}$.

While  (\ref{qf}) is not true for all conceivable $F$'s satisfying
the normalization condition 
\begin{equation}\label{normc}
\int |F|^2\prod_{k=1}^{N}|\pgp(\x_k)|^2d^3\x_k=1 \ ,
\end{equation}
it {\it is} true for an $F$, such as ours, that has bounded
kinetic energy (\ref{bound}). 
In fact, we will show that
\begin{equation}\label{qf3} 
\frac 1N\left[  Q(F) + \eps \sum_{i=1}^{N} \int_{\R^{3N}} |\nabla_i
F|^2\prod_{k=1}^{N}|\pgp(\x_k)|^2d^3\x_k\right] \geq -g
\int|\pgp|^4 - o(1)\ , 
\end{equation} 
for all $F$ satisfying (\ref{normc}), with $\eps=o(1)$ as $N\to\infty$. 

To estimate the left side of (\ref{qf3}) from below, we divide space
into boxes of side length $L$, labeled by $\alpha$, and distribute
the particles over the boxes. If we use Neumann boundary conditions in
each box, this can only lower the energy, since we effectively allow
discontinuous functions. Moreover, since $v$ is positive, we can
neglect interactions among particles in different boxes for the
lower bound. To be precise, the left side of (\ref{qf3}) is bounded
below by
\begin{equation}\label{5.26}
\frac 1N \inf_{\{n_\alpha\}} \sum_\alpha \inf_{F_\alpha} \frac { \widetilde Q_\alpha(F_\alpha) } {\|F_\alpha\|^2} \ , 
\end{equation}
where the infimum is over all distributions of $n_\alpha$ particles in
the boxes $\alpha$, under the constraint that $\sum_\alpha
n_\alpha=N$, and 
\begin{multline}\label{ener3}
\widetilde Q_\alpha(F_\alpha)=\sum_{i=1}^{n_\alpha}\left[ 
\int_{\alpha\cap\Omega_i^c} |\nabla_i
F_\alpha|^2 + \eps \int_{\alpha} |\nabla_i
F_\alpha|^2  \right] \prod_{k=1}^{n_\alpha}|\pgp(\x_k)|^2d^3\x_k\\ 
+\sum_{i=1}^{n_\al} \int_\alpha \left[
\sum_{j\neq i}v(\x_i-\x_j)-2g |\pgp(\x_i)|^2\right] |F_\alpha|^2
\prod_{k=1}^{n_\al}|\pgp(\x_k)|^2d^3\x_k \ .
\end{multline}
Here all the integrals are restricted to the box $\alpha$, and
$F_\alpha$ is a function of $n_\alpha$ variables.

We now fix some $M>0$, that will eventually tend to $\infty$, and
restrict ourselves to boxes $\al$ inside a cube $\Lambda_M$ of side
length $M$. Since $v\geq 0$ the contribution to
(\ref{5.26}) of boxes outside this cube is easily estimated from below by 
$-2gN \sup_{\x\notin \Lambda_M}|\pgp(\x)|^2$, which, divided by $N$,
is arbitrarily small for $M$ large, since $\pgp$ decreases faster than
exponentially at infinity (\cite{LSY00}, Lemma A.5).

For the boxes inside the cube $\Lambda_M$ we want to use the results
on the homogeneous Bose gas obtained in \cite{LY98}, and therefore we
must approximate $\pgp$ by constants in each box. Let $\rmax$
and $\rmin$, respectively, denote the maximal and minimal values of
$|\pgp|^2$ in box $\al$. Define
\begin{equation}
\Psi_\alpha(\x_1,\dots, \x_{n_\al})=F_\alpha(\x_1,\dots, \x_{n_\al}) 
\prod_{k=1}^{n_\al}\phi^{\rm GP}(\x_k)\ ,
\end{equation}
and
\begin{equation}
\Psi^{(i)}_\alpha(\x_1,\dots, \x_{n_\al})=F_\alpha(\x_1,\dots, \x_{n_\al}) 
\prod_{\substack{k=1 \\ k\neq 
i}}^{n_\al}\phi^{\rm GP}(\x_k)\ .
\end{equation}
We have, for all $1\leq i\leq n_\al$,
\begin{multline}\label{5.29}
\left[ \int_{\alpha\cap\Omega_i^c} |\nabla_i
F_\alpha|^2 
 +\half \sum_{j\neq i} \int_\alpha
v(\x_i-\x_j)|F_\alpha|^2 \right] \prod_{k=1}^{n_\al}|\pgp(\x_k)|^2d^3\x_k
\\ \geq
\rmin \left[
\int_{\alpha\cap\Omega_i^c} |\nabla_i
\Psi_\alpha^{(i)}|^2 
 +\half \sum_{j\neq i} \int_\alpha
v(\x_i-\x_j)|\Psi_\alpha^{(i)}|^2 \right] \prod_{k=1}^{n_\al}d^3\x_k \ .
\end{multline}
We now use the following Lemma, which was proved in \cite{LY98}. 
It allows us to replace $v$ by a `soft' potential,
at the cost of sacrificing kinetic energy and increasing the
effective range.

\begin{lem}\label{dysonl} 
Let $v(\x)\geq 0$ with finite range $R_{0}$. Let
$U(r)\geq 0$
be any function satisfying $\int U(r)r^2dr\leq 1$ and $U(r)=0$ for $r<R_{0}$.
Let ${\mathcal B}\subset \R^3$ be star shaped with respect to $0$ (e.g.\
convex with $0\in{\mathcal B}$). Then for all functions $\psi$
\begin{equation}\label{dysonlemma}
    \int_{\mathcal B}\big[|\nabla\psi|^2+\half
v|\psi|^2\big]
\geq  a \int_{\mathcal B} U|\psi|^2 \ .
\end{equation}
\end{lem}

By dividing $\alpha$ for given points $\x_{1},\dots,\x_{n_\alpha}$
into Voronoi cells ${\mathcal B}_{i}$ that contain all points closer
to $\x_{i}$ than to $\x_{j}$ with $j\neq i$ (these cells are star
shaped w.r.t. $\x_{i}$, indeed convex), and choosing a $U$ with radius $R\leq N^{-7/17}$,  we see that (\ref{5.29}) is
bounded below by
\begin{equation}\label{5.30}
(\ref{5.29})\geq  a \rmin\int_\alpha  
U(t_i)|\Psi^{(i)}_\alpha|^2 \geq  a \frac \rmin\rmax \int_\alpha  
U(t_i)|\Psi_\alpha|^2 \ ,
\end{equation}
where $t_{i}$ is the distance of $\x_{i}$ to its nearest
neighbor among the other points $\x_{j}$, $j=1,\dots, n_\al$, i.e.,
\begin{equation}\label{2.29}
t_{i}(\x_{1},\dots,\x_{n_\al})=\min_{j,\,j\neq
i}|\x_{i}-\x_{j}| \ .
\end{equation}
As in \cite{LY98} we choose for $U$ 
the potential
\begin{equation}\label{softened}
U(r)=\begin{cases}3(R^3-R_{0}^3)^{-1}&\text{for
$R_{0}<r<R$ }\\
0&\text{otherwise \ ,}
\end{cases}
\end{equation}
with $R$ determined by $\Omega_\X$ as $R=N^{-7/17}$. Note that $R\gg
R_0$ for $N$ large enough, since $R_0=O(N^{-1})$ in the limit
considered.

Since $\Psi_\alpha=\pgp(\x_i)\Psi^{(i)}_\al$ we can estimate
\begin{equation}\label{tpsial}
|\nabla_i\Psi_\alpha|^2 \leq 2\rmax |\nabla_i\Psi^{(i)}_\alpha|^2 +
2|\Psi^{(i)}_\alpha|^2 C_M
\end{equation}
with
\begin{equation}
C_M=\sup_{\x\in\Lambda_M}|\nabla\pgp(\x)|^2\ .  
\end{equation} 
Inserting (\ref{tpsial}) into (\ref{ener3}), summing over $i$ and
using $\rho^{\rm GP}(\x_i)\leq \rmax$ in the last term of
(\ref{ener3}), we get
\begin{equation}\label{qalfal}
\frac{\widetilde Q_\al(F_\alpha)}{\|F_\alpha\|^2} \geq 
\frac{\rmin}{\rmax}E^{U}_\eps(n_\al,L)-2g\rmax n_\al-
\eps C_M n_\al \ ,
\end{equation}
where $E^{U}_\eps(n_\al,L)$ is
the ground state energy of
\begin{equation}\label{eueps}
\sum_{i=1}^{n_\al}\big(-\half\eps \Delta_i+aU(t_i)\big) 
\end{equation}
in a box of side length $L$. We want to minimize (\ref{qalfal})
with respect to $n_\al$, and drop the subsidiary condition
$\sum_\al{n_\al}=N$ in (\ref{5.26}). This can only lower the minimum.
For the time being we also ignore the last term in
(\ref{qalfal}). The total contribution of this term for all boxes is
bounded by $\eps C_M N$ and will turn out to be negligible compared to
the other terms.

It was shown in \cite{LY98} (see also \cite{lssy}) that
\begin{equation}\label{basicx}
E^{U}_\eps(n_\al,L)\geq \frac{4\pi
an_\al^2}{L^3}(1-CY_\al^{1/17})
\end{equation} 
with $Y_\al=a^3n_\al/L^3$, provided $\eps\geq Y_\al^{1/17}$ and
$n_\al\geq {\rm (const.)} Y_\al^{-1/17}$. The condition on $\eps$ is
certainly fulfilled if we choose $\eps=Y^{1/17}$ with $Y=a^3N/L^3$. We
now want to show that the $n_\alpha$ minimizing the right side of
(\ref{qalfal}) is large enough for (\ref{basicx}) to apply.

If the minimum of the right side of (\ref{qalfal}) (without the last
term) is taken for some $\bar n_\al$, we have
\begin{equation}\label{minnal}
\frac{\rmin}{\rmax}
\left(E^{U}_\eps(\bar n_\al+1,L)-E^{U}_\eps(\bar n_\al,L)\right)\geq 
2g\rmax\ .
\end{equation}
On the other hand, we  claim that
\begin{lem} For any $n$ 
\begin{equation}\label{chempot}
E^{U}_\eps( n+1,L)-E^{U}_\eps(n,L)\leq 8\pi a\frac{ 
n}{L^3}\ .
\end{equation}
\end{lem}
\begin{proof}
Denote the operator (\ref{eueps}) by $\widetilde H_n$, with
$n_\alpha=n$, and let $\widetilde
\Psi_n$ 
be its ground state. Let $t_i'$ be the distance to the nearest
neighbor of $\x_i$ among the $n+1$ points $\x_1,\dots,\x_{n+1}$
(without $\x_i$) and $t_i$ the corresponding distance excluding
$\x_{n+1}$. Obviously, for $1\leq i\leq n$,
\begin{equation}
U(t_i')\leq U(t_i)+U(|\x_i-\x_{n+1}|)
\end{equation}
and
\begin{equation}
U(t_{n+1}')\leq \sum_{i=1}^nU(|\x_i-\x_{n+1}|)\ .
\end{equation}
Therefore
\begin{equation}
\widetilde H_{n+1}\leq \widetilde H_{n}
-\half\eps\Delta_{n+1}+2a\sum_{i=1}^nU(|\x_i-\x_{n+1}|)\ .
\end{equation}
Using $\widetilde\Psi_n/L^{3/2}$ as trial function for $\widetilde
H_{n+1}$ we arrive at
(\ref{chempot}).
\end{proof}
Eq.\ (\ref{chempot}) together with (\ref{minnal}) shows that $\bar
n_\al$ is at least $\sim N \rmax L^3$ (recall that $g=4\pi Na$).  We
shall choose $L\sim N^{-1/10}$, so the conditions needed for
(\ref{basicx}) are fulfilled for $N$ large enough, since $\rmax=O(1)$
and hence $\bar n_\al\sim N^{7/10}$ and $Y_\al\sim N^{-2}$.

In order to obtain a lower bound on (\ref{qalfal}), we can use
$Y_\alpha\leq Y$ in the error term in (\ref{basicx}). We therefore
have to minimize
\begin{equation}\label{qalpha}
 4\pi 
a\left(\frac{\rmin}{\rmax}\frac{n_\al^2}{L^3}\left(1-CY^{1/17}\right)
-2n_\al N \rmax\right) \ , 
\end{equation}
and we can drop the 
requirement that $n_\al$ has to 
be an integer. The minimum of (\ref{qalpha}) is obtained for
\begin{equation}
n_\al= \frac{\rmax^2}{\rmin}\frac{N L^3}{(1-CY^{1/17})} \ .
\end{equation}
Using (\ref{5.26}), (\ref{qalfal}) and (\ref{basicx}) this gives the
following lower bound on the left side of (\ref{qf3}), including now
the last term in (\ref{qalfal}) as well as the contributions from the
boxes outside $\Lambda_M$:
\begin{equation}\label{almostthere}
-g\sum_{\al\subset\Lambda_M} 
\rmin^2
L^3\left(\frac{\rmax^3}{\rmin^3}\frac{1}{(1-CY^{1/17})}\right) -2 g
\sup_{\x\notin\Lambda_M}|\pgp(\x)|^2 -\eps C_M \ .
\end{equation}
Now $\pgp$ is differentiable and strictly positive. Since all the
boxes are in the fixed cube $\Lambda_M$ there are constants
$C'<\infty$, $C''>0$, such that
\begin{equation}
\rmax-\rmin\leq C'L,\quad \rmin\geq C'' \ .
\end{equation}
Since $L\sim N^{-1/10}$ and $Y\sim N^{-17/10}$ we therefore have, for
large $N$,
\begin{equation}
\frac{\rmax^3}{\rmin^3}\frac{1}{(1-CY^{1/17})}\leq 
1+{\rm (const.)}N^{-1/10} \ .
\end{equation}
Also,
\begin{equation}
g\sum_{\al\subset\Lambda_M} \rmin^2 L^3\leq g\int 
|\pgp|^4 \ .
\end{equation}
Hence, for some constant depending only on $g$ and $M$, 
\begin{equation}\label{there}
(\ref{almostthere}) \geq -g\int |\pgp|^4 -{\rm (const.)}N^{-1/10}-2g
\sup_{\x\notin \Lambda_M}|\pgp(\x)|^2 \ .
\end{equation}
This proves the desired result, and finishes the proof of Lemma \ref{L1}.  
\end{proof}

In the following, $\K\subset\R^m$ denotes a bounded and connected set
that is sufficiently nice so that the Poincar\'e-Sobolev inequality
(see \cite[Thm.~8.12]{LL}) holds on $\K$. In particular, this is the
case if $\K$ satisfies the cone property \cite{LL} (e.g., if $\K$ is a
ball or a cube). The following Lemma is a generalization of the
Poincar{\'e} inequality. It can be further generalized to the $L^p$
case, and, with a different and more complicated proof, to the case of
magnetic fields \cite{lsy02}.

\begin{lem}[{\bf Generalized Poincar{\'e} Inequality}]\label{L2} 
For $m\geq 2$ let $\K\subset\R^m$ be as explained above, and let $h$
be a bounded function with $\int_\K h=1$. There exists a constant $C$
(depending only on $\K$ and $h$) such that for all sets
$\Omega\subset\K$ and all $f\in H^1(\K)$ with $\int_\K f h\, d^m\x=0$,
the inequality
\begin{equation} \label{poinc}
\int_\Omega |\nabla f(\x)|^2 d^m\x
+\left(\frac{|\Omega^c|}{|\K|}\right)^{2/m}\int_\K |\nabla
f(\x)|^2 d^m\x \geq \frac 1 C \int_{\K} |f(\x)|^2 d^m\x
\end{equation}
holds. Here $|\cdot|$ is the volume of a set, and
$\Omega^c=\K\setminus\Omega$.
\end{lem}

\begin{proof}
By the usual Poincar\'e-Sobolev inequality on $\K$
(see \cite[Thm.~8.12]{LL}), 
\begin{equation}
\|f\|_{L^2(\K)}^2\leq \widetilde C \|\nabla
f\|_{L^{2m/(m+2)}(\K)}^2
\end{equation}
for some constant $\widetilde C$, if $m\geq 2$ and $\int_\K f h=0$. Using the triangle inequality we can estimate
\begin{equation}
\|f\|_{L^2(\K)}^2 \leq 2\widetilde
C\left(\|\nabla f\|_{L^{2m/(m+2)}(\Omega)}^2+\|\nabla
f\|_{L^{2m/(m+2)}(\Omega^c)}^2\right) \ .
\end{equation} 
Applying H\"older's inequality 
\begin{equation}
 \|\nabla
f\|_{L^{2m/(m+2)}(\Omega)} \leq \|\nabla
f\|_{L^{2}(\Omega)}|\Omega|^{1/m}
\end{equation}
(and the analogue with
$\Omega$ replaced by $\Omega^c$), we see that (\ref{poinc}) holds
with $C=2|\K|^{2/m}\widetilde C$. 
\end{proof}

The important point in Lemma \ref{L2} is that there is no restriction
on $\Omega$ concerning regularity or connectivity. Combining the
results of Lemmas \ref{L1} and \ref{L2}, we now are able to prove
Theorem \ref{T1}.

\begin{proof}[Proof of Theorem \ref{T1}] 
For some $M>0$ let $\K=\{\x\in\R^3,
|\x|\leq M\}$, and define 
\begin{equation} 
\langle f_\X\rangle_\K=\frac
1{\int_\K |\pgp(\x)|^2 d^3\x} \int_\K |\pgp(\x)|^2 f_\X(\x)\, d^3\x \  .
\end{equation} 
We shall use Lemma \ref{L2}, with $m=3$,
$h(\x)=|\pgp(\x)|^2/\int_\K|\pgp|^2$, $\Omega=\Omega_\X\cap\K$ and
$f(\x)= f_\X(\x)-\langle f_\X \rangle_\K$ (see (\ref{defomega})
and (\ref{deff})). Since $\pgp$ is bounded on $\K$ above and below
by some positive constants, this Lemma also holds (with a
different constant $C'$) with $d^3\x$ replaced by $|\pgp(\x)|^2d^3\x$
in (\ref{poinc}). Therefore,
\begin{multline}\label{21}
\int d\X \int_\K d^3\x |\pgp(\x)|^2 \big[f_\X(\x)-\langle
f_\X\rangle_\K\big]^2
\\ \leq C'\int d\X\left[\int_{\Omega_\X\cap \K}
|\pgp(\x)|^2|\nabla_{\x} f_\X(\x)|^2 d^3\x\right. \\ \left.
\qquad\quad\qquad + \frac {N^{-8/51}}{M^2} \int_\K
|\pgp(\x)|^2|\nabla_{\x} f_\X(\x)|^2 d^3\x \right] \ ,
\end{multline}
where we used that $|\Omega_\X^c\cap\K|\leq (4\pi/3)
N^{-4/17}$. The first integral on the right side of (\ref{21})
tends to zero as $N\to\infty$ by Lemma \ref{L1}, and the second is
bounded by (\ref{bound}). We conclude that
\begin{equation}\label{neweq}
\lim_{N\to\infty} \int d\X \int_\K d^3\x |\pgp(\x)|^2 \big[f_\X(\x)-\langle
f_\X\rangle_\K\big]^2 = 0 \ .
\end{equation}
Moreover, since 
\begin{equation}
\int_\K |\pgp(\x)|^2
f_\X(\x) d^3\x\leq \int_{\R^3} |\pgp(\x)|^2 f_\X(\x)d^3\x
\end{equation}
by the positivity of $f_\X$, 
\begin{equation}
\frac 1N \langle
\pgp|\gamma|\pgp\rangle\geq \left[\int_\K
|\pgp(\x)|^2 d^3\x\right]^2 \int d\X  \langle
f_\X\rangle_\K^2 \ .
\end{equation}
Hence, by (\ref{neweq}), 
\begin{equation}
\liminf_{N\to\infty} \frac 1N \langle
\pgp|\gamma|\pgp\rangle\geq \int_\K
|\pgp(\x)|^2 d^3\x \, \lim_{N\to\infty}\int d\X \int_\K d^3\x
|\Psi(\x,\X)|^2 \  .
\end{equation}
It  follows from (\ref{part1}) that the right side of this inequality
equals $\left[\int_\K |\pgp(\x)|^2 d^3\x\right]^2$.  Since the radius
of $\K$ was arbitrary, we conclude that
\begin{equation}
\lim_{N\to\infty}\frac 1 N \langle\pgp|\gamma|\pgp\rangle= 1 \ ,
\end{equation}
implying convergence of $\gamma/N$ to $P^{\rm GP}$ in Hilbert-Schmidt
norm. Since the traces are equal, convergence even holds in trace class norm
(cf. \cite[Thm.~2.20]{S79}), and Theorem~\ref{T1} is proven.
\end{proof}

Throughout the paper we were dealing with Bosons in three-dimensional space. However, the method presented here also works in the case of a
2D Bose gas. The relevant parameter to be kept fixed in the GP limit
is $g=4\pi N/|\ln (a^2 N)|$, all other considerations carry over
without essential change, using the results in \cite{LSY01,LY01}. We
also point out that our method necessarily fails for the 1D Bose gas,
where there is no BEC in the ground state \cite{PS91}. An analogue of
Lemma 1 cannot hold in the 1D case since even a hard core potential
with arbitrarily small range produces an interaction energy that is
not localized on scales smaller than the mean particle spacing.

\bibliographystyle{amsalpha}

\end{document}